\documentclass[journal]{IEEEtran}

\usepackage{cite}
\usepackage{tikz}
\usepackage{amsmath,amssymb}

\usepackage{bm}
\usepackage{graphicx}
\graphicspath{{Figures/}}
\usepackage{subfigure}

\begin{document}
	
\title{Implementation and Characterization of a Two-Dimensional Printed Circuit Dynamic Metasurface Aperture for Computational Microwave Imaging}

\author{Timothy Sleasman, Mohammadreza F. Imani, Aaron V. Diebold, Michael Boyarsky, Kenneth P. Trofatter, and David R. Smith,~\IEEEmembership{Senior Member,~IEEE}.

\thanks{T. Sleasman, M. F. Imani, A. V. Diebold, M. Boyarsky, K. P. Trofatter, D. R. Smith are with the Department of Electrical and Computer Engineering, Duke University, Durham, NC 27708, USA.\newline
Corresponding author: sleasmant@gmail.com}}

\maketitle


\begin{abstract}
We present the design, fabrication, and experimental characterization of a two-dimensional, dynamically tuned, metasurface aperture, emphasizing its potential performance in computational imaging applications. The dynamic metasurface aperture (DMA) consists of an irregular, planar cavity that feeds a multitude of tunable metamaterial elements, all fabricated in a compact, multilayer printed circuit board process. The design considerations for the metamaterial element as a tunable radiator, the associated biasing circuitry, as well as cavity parameters are examined and discussed. A sensing matrix can be constructed from the measured transmit patterns, the singular value spectrum of which provides insight into the information capacity of the apertures. We investigate the singular value spectra of the sensing matrix over a variety of operating parameters, such as the number of metamaterial elements, number of masks, and number of radiating elements. After optimizing over these key parameters, we demonstrate computational microwave imaging of simple test objects.
\end{abstract}


\section{Introduction}

\IEEEPARstart{M}{icrowave} imaging is at the center of a wide swath of applications such as security screening, through-wall imaging, and terrestrial and space observation \cite{Sheen2001, ulaby1981microwave, Ahmed2011, Ahmed2012, Dehmollaian2008}. Systems employed in these scenarios are usually composed of a scanned antenna \cite{skolnik1970radar, Way1991, soumekh1999synthetic, zamani20181} or an array of independently-fed antennas \cite{Ahmed2011, Moulder2016, Gonzalez-Valdes2014, Ghasr2012, Liang2015}. More recently, there has been interest in enlarging microwave imaging systems for improved reconstruction quality and classification performance. The traditional hardware of microwave imaging, i.e. scanned or arrayed antennas, does not provide an affordable and simple scaling path. Instead, recent works have sought to solve this scalability problem by leveraging computational imaging principles \cite{Brady2009a, SOFI, Brady2009, Fergus2006, Sun2013, Liutkus2014, mait2018computational}. 

Computational imaging can be described as a paradigm in which the hardware and software of an imaging system are co-designed with the aim to increase imaging speed, reduce system complexity, or promote novel functionality. Computational imaging emerged as early as the introduction of digital processing techniques in optical imaging \cite{Fenimore1978}. The key notion underlying this increasingly popular trend is to remove the isomorphism between the imaging domain and the hardware sensing layer. In doing so the overall system architecture can be simple and inexpensive \cite{antipa2018diffusercam}, or the data acquisition time can be reduced significantly \cite{Watts2014}. At optical frequencies computational imaging can be achieved by inserting a coded aperture \cite{gehm2006static} or a spatial light modulator \cite{Watts2014, shapiro2008computational} in the signal path. Following the data acquisition, the effects of these encoding devices are taken into consideration in computational post-processing to reconstruct the image.

\begin{figure*}[htbp]
	\centering
	\includegraphics[width=0.78\textwidth]{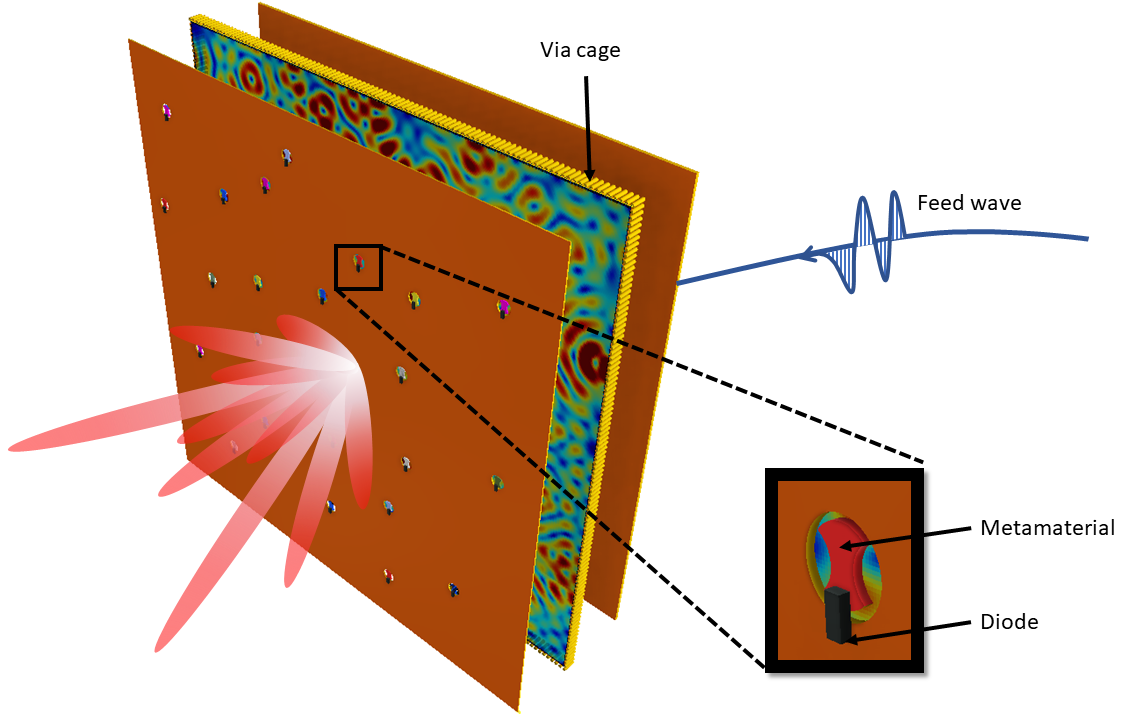}
	\caption{\label{fig:PC_concept}A conceptual schematic of a printed 2D DMA, comprised of a parallel plate converted into a cavity with a via wall. A feed excites the cavity and the elements with spatially-varying magnitudes (highlighted by the colors of the elements) to create a series of radiation patterns.}
\end{figure*}

In microwave imaging systems, metasurface antennas have been demonstrated as useful components for implementing computational imaging schemes. Waveguide-fed metasurface apertures consist of a waveguide exciting an array of metamaterial radiators, each of which has a resonance frequency determined by its geometry and local environment. In one implementation of a metasurface aperture, the resonant frequencies of the metamaterial elements are selected to be randomly distributed across a band of operation \cite{Hunt2013, Hunt2014, Lipworth2013}. As the frequency of operation is changed, different subsets of elements are excited, producing diverse radiation patterns that encode scene information as a function of driving frequency. The number of independent radiation patterns available in this modality is determined by the overall quality (Q-) factor of the aperture, which can be improved by replacing the waveguide feed with a disordered, electrically-large cavity, even when using non-resonant metamaterial radiators \cite{Fromenteze2015, yurduseven2016multistatic, Yurduseven2016mills, F.Imani2016, Gollub2017}. The frequency-diverse metasurface aperture is capable of producing high-resolution RF images; however, the ultrawide bandwidth of operation required inevitably increases the cost and complexity of the system. An alternative approach is to independently address and tune each of the metamaterial elements, forming a dynamic metasurface aperture (DMA) \cite{Sleasman2016a, Sleasman2015, Sleasman2016Design}. By dynamically changing the aperture distribution of radiating elements, diverse radiation patterns can be generated and used to encode scene information with minimal bandwidth---potentially even a single frequency of operation \cite{sleasman2017single, boyarsky2018single, Diebold, diebold2018optica}.

The design of a two-dimensional (2D) DMA---consisting of a planar, printed cavity loaded with tunable metamaterial elements---was recently proposed and studied in \cite{imani2018two}. Here, we report a physical implementation of a 2D DMA, demonstrating the design of the metamaterial radiator; the biasing circuitry used for the independent tuning of each element; the cavity feed; and the element distribution. The fabricated sample is then experimentally characterized and important factors governing its imaging performance are identified. We provide a thorough characterization of the fabricated sample's reflection coefficient and its relationship to the cavity parameters. From experimental near-field scans of the aperture, we are able to determine the far-field radiation patterns generated by the aperture both as a function of frequency as well as tuning state. Knowledge of these radiation patterns can be used to form the sensing matrix, which connects the measured return signal to the conductivity distribution of the scene. By studying the singular value spectra of the sensing matrix for different operating conditions---such as the number of metamaterial elements; the number of radiating versus non-radiating elements; and the potential grouping of elements---we can assess the information content and optimize the overall imaging system. After completing this analysis, we demonstrate a small-scale imaging system with the 2D DMA as the transmitter and four open-ended waveguide probes as receivers.

\section{Design of the 2D DMA}

The printed-cavity-backed 2D DMA developed in this paper is be designed to operate at K-band (20-24 GHz). The metamaterial radiators are tuned by the inclusion of PIN-diodes integrated into the metamaterial elements. The metamaterial radiators thus can be tuned to be radiating or non-radiating, which we refer to as \textit{binary} tuning. In the following, we describe the design process for each component of the 2D DMA assuming binary elements.

\subsection{Metamaterial element design}
The first step in our design process is to design metamaterial elements. One advantage of using a metamaterial resonator as a tunable radiator is that its resonant response provides a simple means for switching it \emph{on} or \emph{off} by pushing its resonance frequency out of the operating band. For a metasurface aperture operating at K-band frequencies, each of the metamaterial elements must thus be resonant within the K-band. In addition to being a resonant radiator, the metamaterial element should couple strongly to the feed wave when in the \emph{on} state, since relatively few elements will be included in the final design (to ensure low cost), of which only a subset are radiating (turned \textit{on}) at any given time. An element with high coupling efficiency guarantees that ample energy is radiated into the scene.

\begin{figure}[!t]
	\centering
	\includegraphics[width=\linewidth]{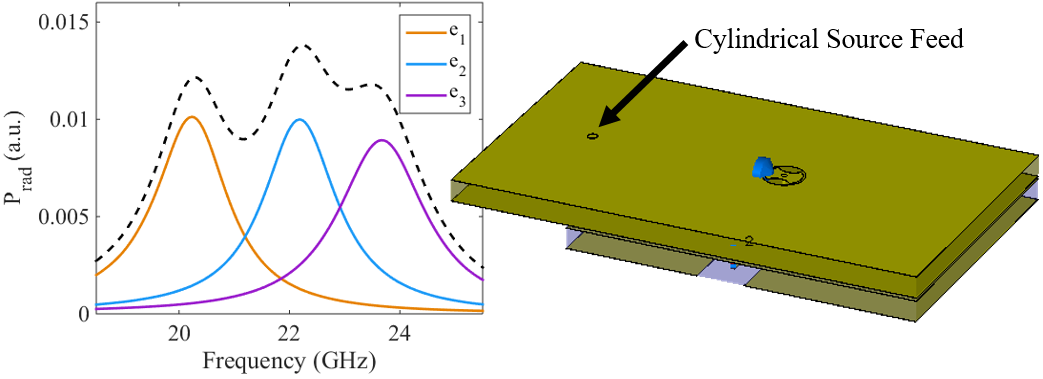}
	\caption[Radiation from cELC Elements]{\label{fig:PC_element}Three separate unit cells are utilized to broaden the bandwidth. The black dashed curve represents the summation among the elements, which gives the spectral power radiated distribution. Full-wave simulations in CST are used to calculate the radiated power from the configuration shown, which includes a model of the element, the diode (blue), and the biasing circuitry (seen below the ground plane).}
\end{figure}

Given the above constraints, we chose the rounded cELC (complementary electric resonator) introduced in \cite{yoo2016efficient} as the metamaterial radiator. This element also exhibits good polarization purity compared to the more common rectilinear cELC \cite{Odabasi2013,Hand2008}. An illustration of the rounded cELC can be seen in Fig. \ref{fig:PC_element} in the simulation setup used in our design process. The element is inserted in a parallel plate waveguide which is fed by a cylindrical source. The simulation domain is terminated with a perfectly matched layer (PML) so that the element only experiences the fields from the driving cylindrical wave.

To arrive at the desired resonance frequency, we need to include all parasitic components affecting the metamaterial element response. The diode is placed across the capacitive gap of the cELC. The central patch of the cELC serves as one contact of the diode, with the top part of the parallel plate waveguide serving as the other. The cELC is connected to a biasing line, which lies below the ground plane, using a via that connects to the central patch. The via is a strong scatterer in the waveguide, but this is not problematic since it serves to further scatter the guided wave (making the cavity modes more diverse) \cite{dupre2014using,F.Imani2016}. The parasitic inductance associated with the via must be included in the simulation to ensure the designed element resonates at the desired frequency. Coupling circuitry is also included in the simulation (below the ground plane). 

\begin{figure}[!t]
	\centering
	\includegraphics[width=\linewidth]{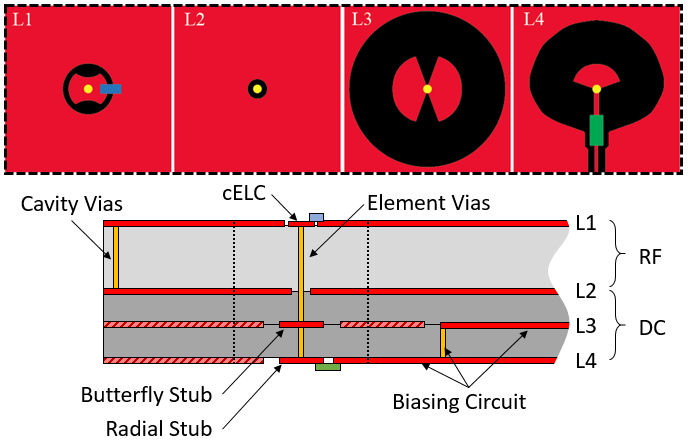}
	\caption[DMA PCB Layers]{\label{fig:PC_layers}Layers of the printed circuit board, with a via from layer 1 through 4 (L1--L4) in yellow, copper pours represented in red, and the diode/resistor shown in blue/green. L1--L2 constitute the parallel plate, waveguide while L3 and L4 are used for bias circuits and decoupling. Some portions of copper (denoted by diagonal lines in the stackup) are for mechanical integrity.}
\end{figure}

On the layer where the via connects to the bias line (below the bottom plate of the waveguide) we insert a radial stub that acts as a low impedance node. This reflects any RF signal traveling on the via back up into the waveguide so that RF energy is not leaked into the biasing circuitry \cite{Sleasman2016a}. Since the DC biasing circuitry requires two layers (making four total layers, as counted by copper-containing planes) we have two layers below the parallel plate waveguide. In addition to the radial stub placed on layer four (L4), a butterfly stub is also included on layer three for the same purpose (L3). The copper planes in L1--L2 thus constitute the parallel plate waveguide while L3--L4 are used for biasing circuitry, as shown in Fig. \ref{fig:PC_layers}. The substrate between L1--L2 is a low loss dielectric for use in RF applications, Rogers 4003, to minimize losses for the signal in the cavity. The substrates below are FR4 since only digital signals exist on these layers.

The PIN diode is modeled in the full wave simulation as a series inductor-resistor circuit in its conducting state (when the element is not radiating) and as an inductor in series with a parallel capacitor-resistor in its non-conducting state (when the element is radiating) \cite{Sleasman2016a}. The inductance, resistance, and capacitance of the element are based on the data available on the datasheet. For our design, we used the MACOM MADP000907-14020W PIN diode.

With all potential parasitic effects and the capacitance of the diode included, we simulated the element in the setup shown in Fig. \ref{fig:PC_element} using CST Microwave Studio. The radiated power of the element as a function of frequency is used as the indication of element's resonant response. To set the resonance frequency of the element to be within the desired range, we have changed the element's proportions. While doing so, we ensured the capacitive gap remained smaller in size than the size of the PIN diode. We design three separate elements---each with a different resonance frequency---to ensure we can operate over a reasonable bandwidth. The diameter of the smallest element is on the order of 3.5 mm, and a scale factor is utilized to create the two slightly larger elements (with lower resonance frequencies). Their simulated radiated powers as a function of frequency are shown in Fig. \ref{fig:PC_element}. The result is reported with arbitrary units since the element only interacts with a portion of the energy fed into the waveguide. We can see from the black dashed line that extends around $20$ to $24$ GHz that the three elements combined cover a large portion of the K-band. For each element, $e_1$--$e_3$, we simulated with the element coupled to the co-pol (as shown) as well as the cross-pol (rotated $90^\circ$) to confirm polarization purity. Each element has also been simulated in its \emph{off} state to confirm that it does not radiate into either the co- or cross-pol.

\subsection{Biasing circuitry}
The biasing circuitry that exist on layer \emph{L4} consists of 8-bit shift registers (to distribute the control), buffers (to ensure enough current is sourced), decoupling capacitors (to stabilize the digital signals), and current-limiting resistors. Each PIN diode draws $<5$ mA when it is forward biased, and negligible current otherwise. This current is supplied by an Arduino microcontroller, which is also responsible for transmitting the digital control signals. All of the diodes (among all of the elements) are connected to the top plate of the waveguide, making it the DC ground. All elements are biased using aforementioned vias.

\subsection{Cavity design}
The last key component of the printed cavity DMA is the via fence that renders the waveguide a reverberating structure. Metallic vias are placed along a perimeter and reflect the outgoing cylindrical wave originating at the coax feed. The result is a standing wave (as shown in Fig. \ref{fig:PC_concept}) which varies with frequency. The via fence has subwavelength spacing and is stacked two layers deep to mitigate leakage. The distance between the two via fences is made small enough that no guided wave (for instance a substrate-integrated-waveguide mode) is supported. We make the cavity perimeter an irregular shape by including an inward-bowing deformity. This assists in mitigating \emph{regular} or \emph{tangential modes}, as defined in \cite{gros2014universal}, and increases the chaotic behavior in the cavity. Combined with the effects of the biasing vias, this irregular shape ensures that the cavity modes are spatially variant and highly frequency dependent and also delivers a uniform excitation across the entire aperture (over many tuning states and frequency points).

\begin{figure*}[!t]
	\centering
	\includegraphics[width=0.75\textwidth]{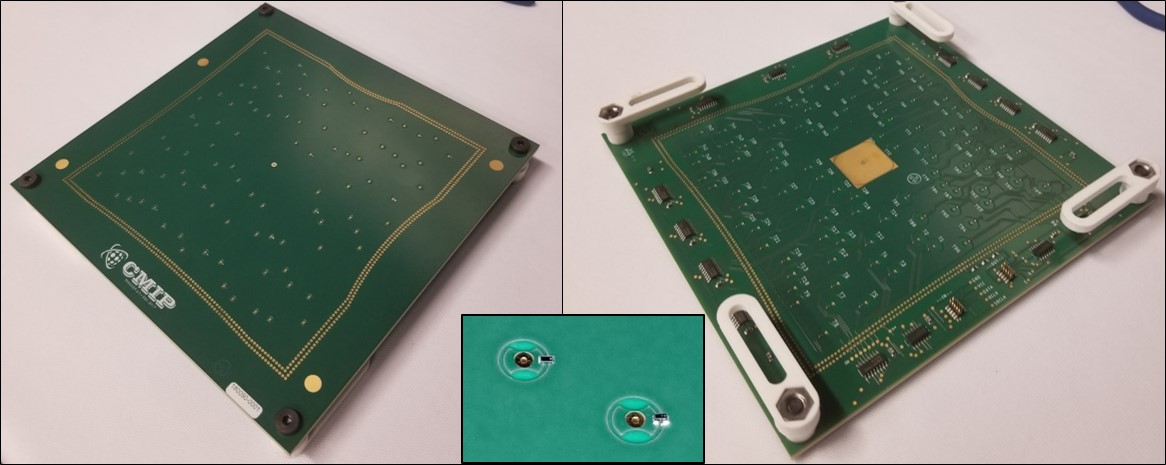}
	\caption[Fabricated DMA]{\label{fig:PC_fabbed}Pictures of the front (left) and back (right) of the fabricated DMA. The inset shows a close up picture of the cELC elements.}
\end{figure*}

\subsection{Aperture size and elements' layout}
 The dimensions of the via fence, and therefore the active area of the aperture, is selected to be around $15{\times}15 \text{ cm}^2$. This size is selected based on a trade-off between dielectric losses, fabrication constraints, and imaging performance (larger aperture size results in higher resolution). The placement of the cELCs is chosen pseudo-randomly within the via fence, with the condition that the elements are not closer than a certain proximity threshold (to ensure tuning circuitry of different elements do not interfere with each other). The random element placement provides fairly uniform coverage in the spatial frequency domain. In other words, for any point in the scene these randomly-placed elements will illuminate the objects with a large collection of effective plane waves. As discussed in \cite{Marks2016}, having this support in the spatial frequency domain (i.e. $k$-space) enables high-quality imaging results. The randomness also reduces aliasing effects that arise from the sparsity of the aperture, which would otherwise result in large sidelobes \cite{Marks2016,Yurduseven2016mills,Yurduseven2016Printed}. 
 
 Figure \ref{fig:PC_fabbed} shows the ultimate implementation with all of the components described above. The semiconductor elements can be seen on the top and bottom layers, and a close up of the rounded cELCs is also shown. In total, $96$ elements are utilized. Out of $96$ elements, we selected $32$ to be of each geometry (see Fig. \ref{fig:PC_element}). Using more elements would increase the cost and complexity of the system. Additionally, including too many elements would require more vias that penetrate through the PPWG, which could create stop bands and undesirable interactions in the waveguide layer. The choice of $96$ elements is therefore practical for a prototype of the printed cavity-backed DMA; the effectiveness of this choice is justified quantitatively in the results that follow.

\section{Experimental Characterization of the DMA}
\label{sec:pc4}

With the printed cavity-backed 2D DMA conceptualized and fabricated, we can now investigate the merits of the proposed structure experimentally. We begin by looking at the structure's reflection coefficient to assess the insertion loss of the device. Near-field scans are then used to experimentally characterization the field patterns generated by the DMA as a function of tuning state and frequency. These scans can be propagated throughout the imaging domain, where we can calculate valuable statistics to help analyze the DMA's performance.

\subsection{Reflected Signal}
\label{ssec:pc4a}

\begin{figure}[!t]
	\centering
	\includegraphics[width=\columnwidth]{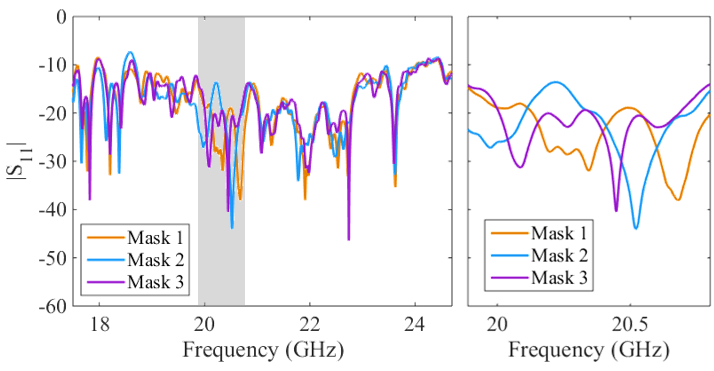}
	\caption[Reflection Coefficient]{\label{fig:PC_s11}The reflection coefficient of the DMA for three different masks. The variation indicates tuning of the elements in the aperture.}
	
	\includegraphics[width=\columnwidth]{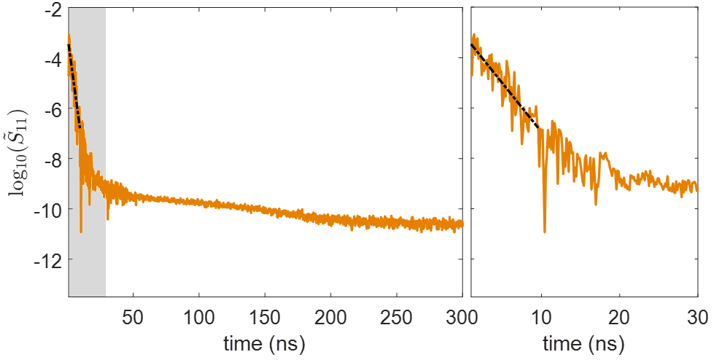}
	\caption[Time-Domain Reflection Coefficient]{\label{fig:PC_s11_td}The time domain response of the printed-cavity for a given mask, from which the $Q$ factor can be deduced.}
\end{figure}

The first experimental characterization of the 2D DMA involves measuring the reflected signal, or \emph{return loss}, at the input port. The measured reflected signal is plotted in Fig. \ref{fig:PC_s11} as a function of frequency. The many dips apparent in this plot denote cavity resonance frequencies where the inserted signal couples into the cavity's eigenmodes. We expect a high density of modes given the cavity's large electrical size. However, there should be some spectral spreading of these modes because of the losses associated with the leaking elements \cite{Dupre2015}. It can be seen that when the elements are tuned into radiating states, creating different masks, the cavity modes are modified (manifested as the variation in the reflection signals dips) so that a new set of modes couple to the connecting port. Because the modes are reconfigured, the elements are excited by new fields for each of these masks. 

\begin{figure}[!t]
	\centering
	\includegraphics[width=\columnwidth]{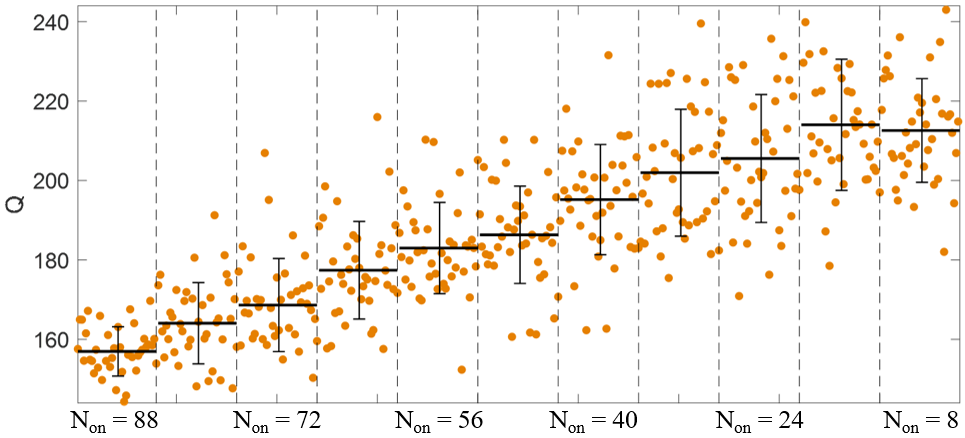}
	\caption[Q for Different Masks]{\label{fig:PC_Q}Quality factors for numerous masks and different \emph{on}/\emph{off} ratios. Notice that $Q$ is reduced when more elements are in their radiating state. The mean and standard deviation are also shown.}
\end{figure}

When plotted in the time domain, the reflected signal can also be used to calculate the quality factor of the structure. This is done by examining the decay of the time-domain impulse response \cite{yurduseven2016multistatic}. The experimental result as obtained from the linear fit in Fig. \ref{fig:PC_Q} is around $200$. This value is the quality factor of the loaded cavity with the impact of the radiation losses. The intrinsic quality factor of this cavity without the radiative layer can be estimated theoretically through the loss tangent of the host dielectric, which in this case is $Q_{dielectric} \approx 1/\tan \delta \approx 370$. Reduction from this value is likely caused by losses in the diodes, ohmic losses, and leakage through the elements. With more of the elements placed in their radiating state, we expect the energy to be leaked from the cavity more quickly, and thus lower loaded quality factor. To test this, we measured the time-domain reflection coefficient and $Q$ factor for a large collection of masks and for different \emph{on}/\emph{off} ratios. For each \emph{on}/\emph{off} ratio we computed the mean and standard deviation over 20 different masks, as plotted in Fig. \ref{fig:PC_Q}. With more elements in the radiating state the $Q$ factor is decreased. The range of measured Q factors was $145$--$245$. If we were operating without dynamic tuning, this result would suggest that we should use a specific number of cELC elements to ensure a high quality factor (and consequently, high frequency diversity). In this sense, our result confirms the theory in \cite{Marks2016} which identifies the trade-off of number of elements, $N_e$, vs. $Q$ as a crucial design criterion.

\subsection{Radiation Patterns}

Though the results from the reflection coefficient study provide insight into the fields in the cavity, illuminating details about the radiation patterns generated by this device can be extracted from its near field scans (NFSs) \cite{Yaghjian1986}. Using the scanned data, we can examine the power radiated by the 2D DMA by summing over all the data. The results of this analysis, which have been omitted for brevity, provide two observations into the operation of the fabricated sample: (1) when more elements are in the \emph{on} state, radiation increases, as expected, and (2) the frequencies between $17.5$--$22$ GHz radiate the most. This red shift (relative to the designed elements, cf. Fig. \ref{fig:PC_element}) is not surprising as it existed in other DMAs, and can be attributed to fabrication and component tolerances. 

\begin{figure}[!t]
	\centering
	\includegraphics[width=\linewidth]{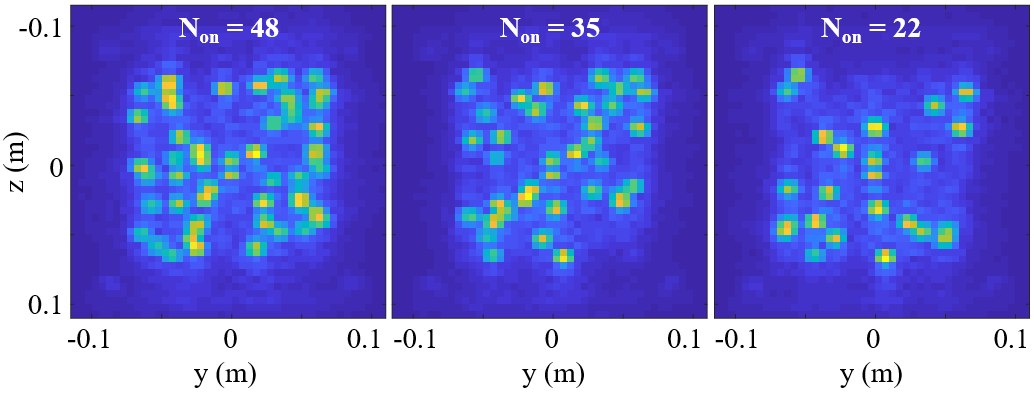}
	\caption[Backprojected Aperture Fields]{\label{fig:PC_backprop}Effective sources in the aperture plane as calculated from backpropagation. The number of bright source magnitudes, averaged over the bandwidth, are in agreement with the number of \emph{on} elements.}
\end{figure}

The red shift in resonance frequencies of the elements suggests that we should operate in the lower portions of K band if high SNR is desired. However, it does not confirm how our collection of elements are tuning since it is completely devoid of spatial information. It is therefore helpful to back-propagate the measured NFS data to the aperture plane to see the field at the aperture of the 2D DMA. We have shown the result of this process for three different masks, with different $N_{on}$, in Fig. \ref{fig:PC_backprop}. Each bright spot in these plots corresponds to a radiating cELC, and it can be seen that the correct number of elements are in their radiating state. We have also completed this study with a single element in the radiating state at a time and observed that every element is functioning as anticipated ($100\%$ yield). Additionally, since the different elements have different power radiated spectra we have confirmed that each element radiates in the expected bandwidth. These studies are omitted for brevity.

\begin{figure*}[htbp]
	\centering
	\includegraphics[width=0.75\textwidth]{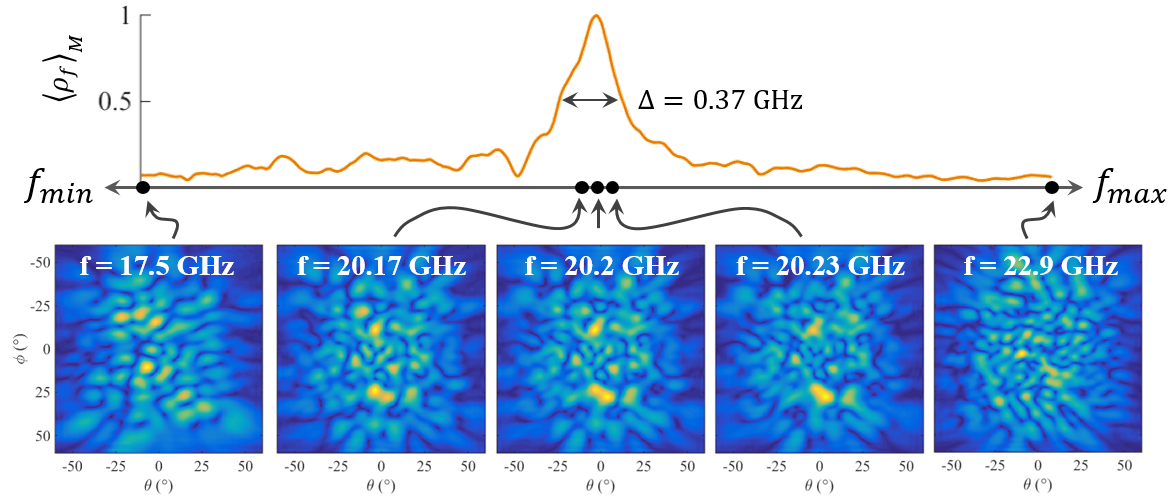}
	\caption[Variation by Changing Frequency]{\label{fig:PC_corr_freq}The dependence of the radiation patterns based on varying frequency. Widely separated frequency points are shown, along with a cluster in the middle (the center three plots). Nearby frequencies exhibit similar radiation patterns, as is illustrated with the correlation plot $\langle\rho_f\rangle_M$ which is averaged over a sequence of $10$ masks.}
	\includegraphics[width=0.75\textwidth]{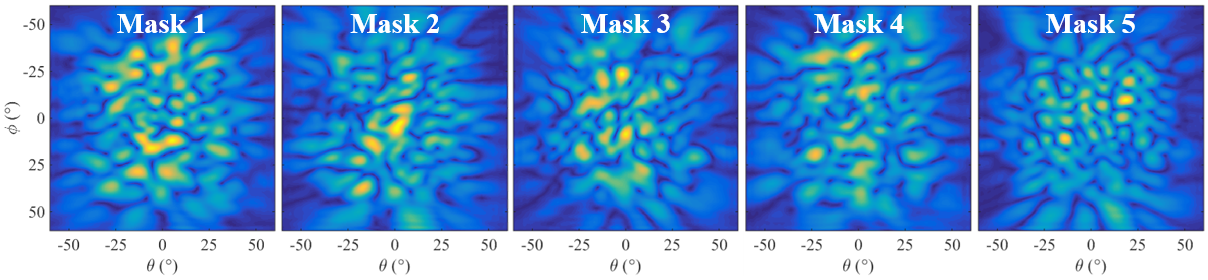}
	\caption[Variation by Changing Masks]{\label{fig:PC_corr_mask}The variation in radiated patterns achieved by tuning through different random masks (shown at 20.2 GHz).}
\end{figure*}

The back-projection results confirm the operation of the fabricated planar DMA. To evaluate imaging performance, we propagate these aperture fields toward the imaging domain. We have plotted projected patterns in Figs. \ref{fig:PC_corr_freq}--\ref{fig:PC_corr_mask} for different frequencies and masks, respectively. As shown, the radiation patterns change with frequency and with mask tuning, and there are different levels of correlation between the radiated patterns among the modulation schemes. To quantify the correlation, we compute a four dimensional function that can be defined as \cite{thaysen2006envelope}

\begin{align}
\label{eq:pc_corr}
&\rho(f_1,f_2,T_1,T_2) = \notag\\
&\frac{\displaystyle \Bigg\lvert \int_S E(f_1,T_1,\overline{r}) E^*(f_2,T_2,\overline{r})  d\overline{r}\Bigg\rvert}
{\displaystyle \sqrt{\int_S |E(f_1,T_1,\overline{r})|^2 d\overline{r}}   \sqrt{\int_S|E(f_2,T_2,\overline{r})|^2 d\overline{r}}}
\end{align}

\noindent where $E$ is the electric field of the DMA at the scene, $f_1$ and $f_2$ are the frequencies of two radiation patterns, $T_1$ and $T_2$ are their tuning states, and $*$ denotes complex conjugation. This equation gives the correlation as a function of frequency/masks over a region of space. As such, the integral is taken over a surface $S$ which we define as a hemispherical shell in the far field spanning $\pm 50 ^\circ$. We can take a subset of this 4D space by looking at the correlation in frequency

\begin{equation}
\label{eq:pc_corr_f}
\rho_f(f) \equiv \rho(f , f_c , m, m), \,\,\,\,\,\,\,\,\,\,\, (f_c \text{ and } m \text{ constant})
\end{equation}

\noindent where we have selected a specific frequency (the center frequency $f_c$) and a fixed mask $m$. So that we are not misdirected by a pathological mask, we can average this over an ensemble $M$, $m \in M$, to get $\langle \rho_f \rangle_M$, where $\langle \cdot \rangle$ denotes averaging, as shown in Fig. \ref{fig:PC_corr_freq}. We can clearly see that we have around $0.37$ GHz of frequency correlation \cite{lerosey2004time}, indicating that patterns at frequencies closer than that this exhibit high correlation as can also been in Fig. \ref{fig:PC_corr_freq}. In other words, relying on frequency diversity alone limits us to very few distinct measurements when operating over a narrow bandwidth.

Turning our attention to the dynamic case we can consider another subset of this 4D space where the masks are tuned (see Fig. \ref{fig:PC_corr_mask}). Since there is no natural order to the mask dimension we cannot gradually vary this parameter. We can, however, fix the frequency and calculate the correlation for two masks, $m$ and $n$, as

\begin{equation}
\label{eq:pc_corr_m}
\rho_{mn}(m,n) \equiv \rho(f_c, f_c , m, n), \,\,\,\,\,\,\,\,\,\,\, (f_c \text{ constant}).
\end{equation}

\noindent We then calculate an average over an ensemble of $m,n \in M$ to obtain $\langle \rho_{mn} \rangle_M = 0.342$. Calculating $\langle \rho_{mn} \rangle_M$ for $f_\text{const.}{\neq}f_c$ returns a similar result across the entire bandwidth. Since the fields in the cavity may not change dramatically when a single frequency is used, we would expect some correlation in the radiation pattern to exist. To decrease the correlation even further, we can change both the mask and frequency simultaneously. As a last correlation calculation, we can therefore define

\begin{equation}
\label{eq:pc_corr_mf}
\rho_{fm}(f,m) \equiv \rho(f, f_c , m, n), \,\,\,\,\,\,\,\,\,\,\, (m \neq n, f_c \text{ constant}).
\end{equation}

\begin{figure}[!t]
	\centering
	\includegraphics[width=0.8\columnwidth]{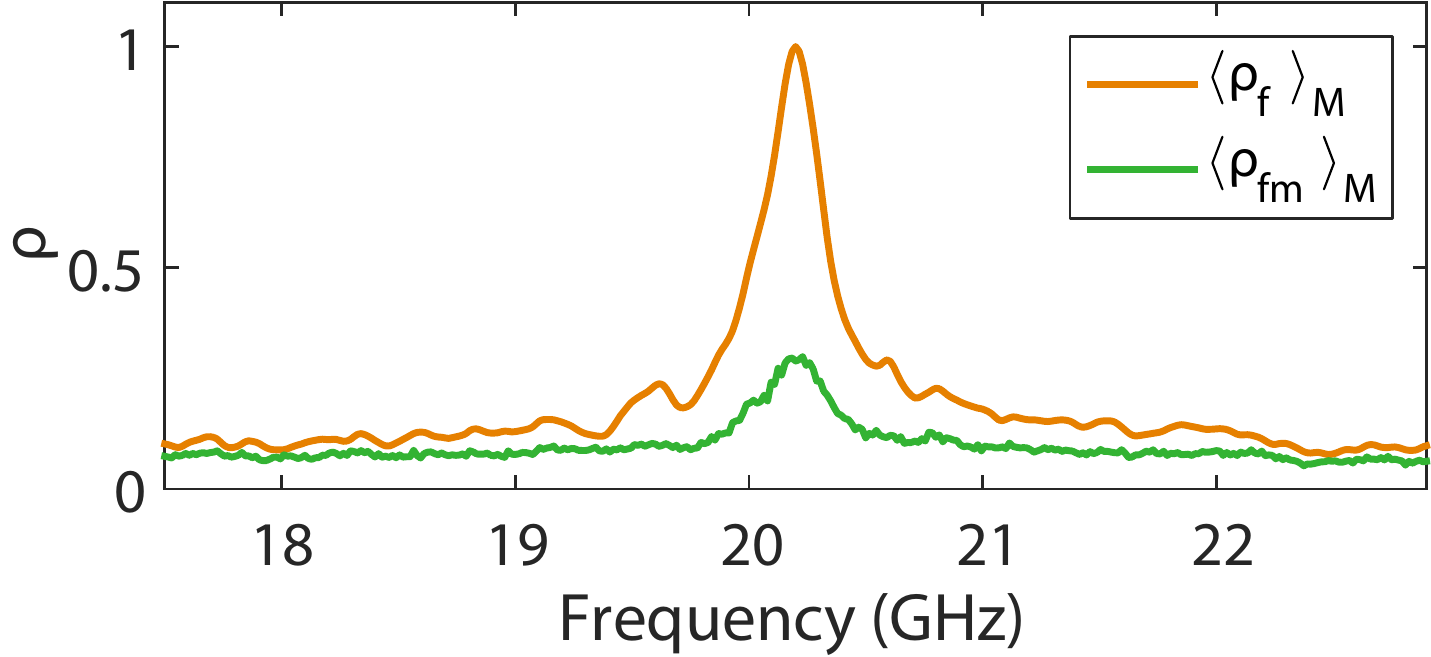}
	\caption[Correlation of Fields]{\label{fig:corr_freq}The correlation statistics with and without dynamic tuning, as described in Eqs. \ref{eq:pc_corr_f} and \ref{eq:pc_corr_mf}.}
\end{figure}

\noindent Again, averaging over numerous instances of $m$ and $n$, we can obtain an ensemble average $\langle \rho_{fm} \rangle_M$, which is plotted in Fig. \ref{fig:corr_freq}. As can be seen, there is a peak around the fixed center frequency, even for the dynamic case, which is caused by the static cavity mode. Nonetheless, there is a significant reduction in the correlation across the bandwidth, which indicates enhanced spatial diversity in the radiated patterns. In calculating this correlation, we are operating as if a new mask is used for each frequency point. In our current implementation, it is slower to change frequency and mask simultaneously, so going forward we will complete a frequency sweep for each mask and include all combinations. 

\subsection{SVS Analyses}
\label{ssec:pc4a_prime}

Calculating correlations is useful for determining the overlap among a collection of masks, but for imaging purposes, it is more helpful to look at the corresponding sensing matrix $H$. We can then take the singular value decomposition of this matrix to examine the the amount of distinct information the DMA can retrieve from the scene \cite{Gollub2017}. To better illustrate this point, we briefly review the role of the sensing matrix in describing the imaging process. In an imaging configuration, the DMA illuminates a scene of interest with patterns described by $\vec{E_T}$. These patterns, scattered by the target, are then collected by a receiver. The signal at the receiver is then an $M\times 1$ vector $\mathbf{g}$ ($M$ is the number of measurements). Assuming the first Born approximation and diffraction limited imaging, the $M\times P$ sensing matrix defined as \cite{Lipworth2013}:

\begin{equation}
    \mathbf{g}=\bm{H\sigma}, \label{eq:main}
\end{equation}

\noindent where $\bm{\sigma}$ is the $P\times 1$ reflectivity map of the scene ($P$ is the number of voxels in the scene), and the sensing matrix is given by

\begin{equation}
    H(m,p)=\vec{E_T^m}(\vec{r_p}).\vec{E_R^m}(\vec{r_p}).
\end{equation}

In the above, $\vec{E_T^m}(\vec{r_p})$ and $\vec{E_R^m}(\vec{r_p})$ are respectively the fields due to the transmitter and receiver, corresponding to $m$th measurement, at the scene location $\vec{r_p}$. Equation (\ref{eq:main}) has to be inverted to find the reflectivity map of the scene. In reality, $\mathbf{H}$ is rarely an invertible square matrix, and computational techniques such as least square solvers are used to estimate $\bm{\sigma}$. In this process, the effective rank of $\mathbf{H}$ is crucial for high fidelity imaging. A common technique to assess the level of correlation is to compute the singular value decomposition of $\mathbf{H}$. A rapidly decaying singular value spectrum indicates a large level of correlation among measurements described by $\mathbf{H}$. In this manner, the slope of an SVS plot can be used to compare different imaging configurations and the role of different features in the imaging process.

\subsubsection{ Frequency-diverse vs. dynamic metasurface}
\begin{figure}[!t]
	\centering
	\includegraphics[width=\columnwidth]{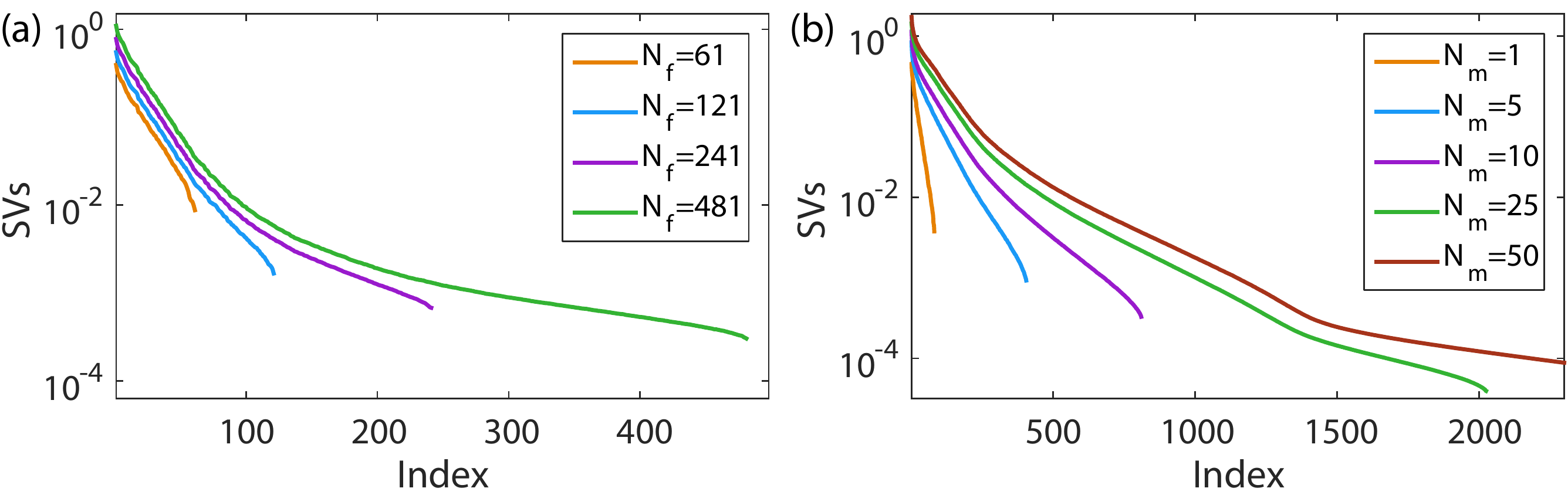}
	\caption[Singular Values for Many Frequencies and Masks]{\label{fig:svd_fm_combo}(a) Singular values when no tuning is used ($N_m=1$). The bandwidth is kept constant and denser frequency sampling is used to increase $N_f$. (b) Singular values when tuning is used and for an increasing number of masks. A constant bandwidth is again used, with $N_f=81$.}
\end{figure}

The fabricated sample provide a unique opportunity to directly compare the expected performance of a broadband, frequency-diverse metasurface aperture with that of a narrow band, dynamic one. To focus on the role of different parameters in retrieving cross range information (and to avoid calculating the SVD of large matrices) we will complete this process on a hemispherical shell in the far field, creating $\mathbf{H}_\text{farfield}$ (similar to the studies in \cite{imani2018two}). We assume the receiver is an open ended waveguide which is isotropic and does not contribute to spatial resolution. As with the correlation studies, we can begin by looking at the singular values based on frequency diversity, as shown in Fig. \ref{fig:svd_fm_combo}a. A bandwidth of $B=7.2$ GHz is used ($17.5$--$24.7$ GHz) with a variable number of frequency points. According to \cite{Marks2016} and the quality factors from Fig. \ref{fig:PC_Q}, the number of useful frequency points should be approximately $85$, with diminishing returns thereafter. From Fig. \ref{fig:svd_fm_combo}a the SVS begin to gradually saturate around the $100$th--$150$th point. Improvements are still seen as the number of frequency points is increased, but there is more correlation among the fields. We have also plotted the singular values for the dynamic case in Fig. \ref{fig:svd_fm_combo}b for random tuning states with $50\%$ of the elements in their \emph{on} state. In this plot, the same bandwidth is used and we show the enhancements possible with dynamic capabilities and an increasing number of masks $N_m$. Note that the $N_m=1$ case corresponds to purely frequency-diverse operation.

\begin{figure}[!t]
	\centering
	\includegraphics[width=\columnwidth]{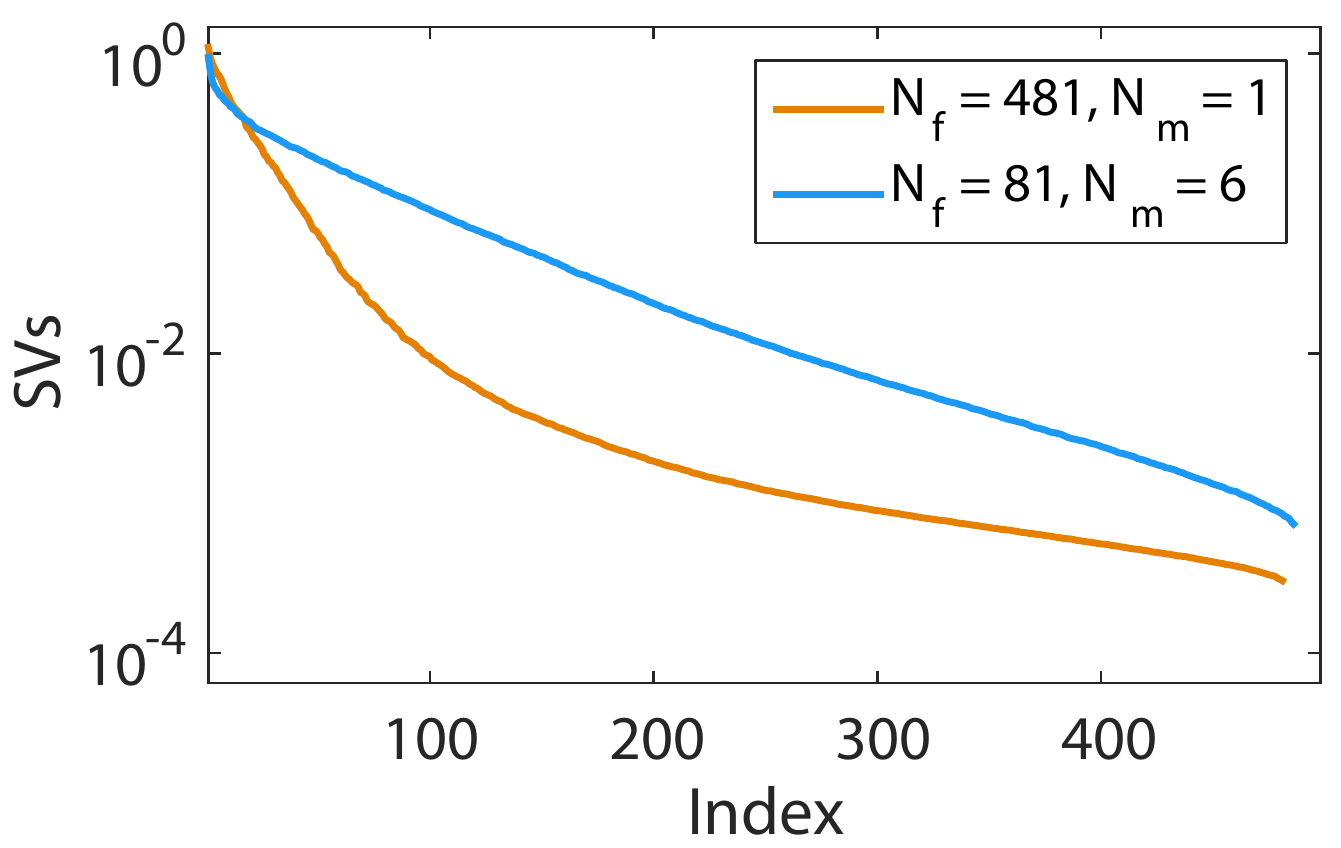}
	\caption[Singular Values Combining Frequency and Masks]{\label{fig:svd_fm}SVS of the passive and dynamic cases, with approximately the same number of measurements.}
\end{figure}

The results in Fig. \ref{fig:svd_fm_combo} show how information scales when an increasing number of measurements is used, but a more useful and fair comparison between passive and dynamic systems can be made when the same number of overall measurements is used. To make this comparison, we use a denser frequency vector across the same bandwidth to increase the number of frequency-diverse measurements. For the passive case we use $481$ frequency points; for the dynamic case we use $81$ frequencies and $6$ masks (random tuning states with $50\%$ of the elements in their \emph{on} state). The results for both cases are shown in Fig. \ref{fig:svd_fm}. The dynamic case has a flatter slope, indicating that the system can obtain more information. If more masks are used, as seen in Fig. \ref{fig:svd_fm_combo}b, the dynamic case will only become better. Notice the similar behavior between the results in this plot, for the printed DMA, and the tunable disordered cavity used in \cite{Sleasman2016disorder}.

\subsubsection{Maximum number of masks}

\begin{figure}[!t]
	\centering
	\includegraphics[width=0.8\columnwidth]{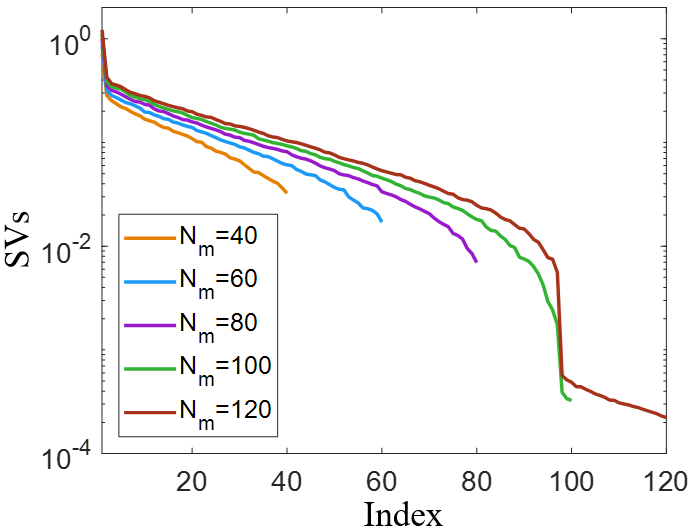}
	\caption[SVS: Varying $N_m$]{\label{fig:PC_svd_f1_nmasks}The behavior of the singular value spectra with an increasing number of masks, $N_m$.}
\end{figure}

Before assessing how different tuning mechanisms perform in an imaging setting, it is important to understand the theoretical limitations on the useful number of masks, $N_m$. For a given frequency, we can continually increase the number of masks until the SVS reaches a drop off. This is shown in Fig. \ref{fig:PC_svd_f1_nmasks} where it is seen that there is a sharp cut-off at $N_m=96$, which is equal to the number of elements $N_e=96$. This is expected since the radiation patterns are linear combinations of the contributions from the independent elements. Put another way: turning on $e_1$ and then $e_2$ gives the same information as turning on $e_1+e_2$ and then $e_1$ (and then subtracting the two). When operating with complete control over each element, this is the limitation on the useful number of measurements available to the aperture at a single frequency. Note that $N_m=96$ is smaller than the $\text{SBP}_\text{aperture}$ (${\approx}400$) since the elements have spacing $>\lambda/2$ \cite{Lohmann1996}.\\

\subsubsection{Number of \emph{on} elements}

\begin{figure}[!t]
	\centering
	\includegraphics[width=\columnwidth]{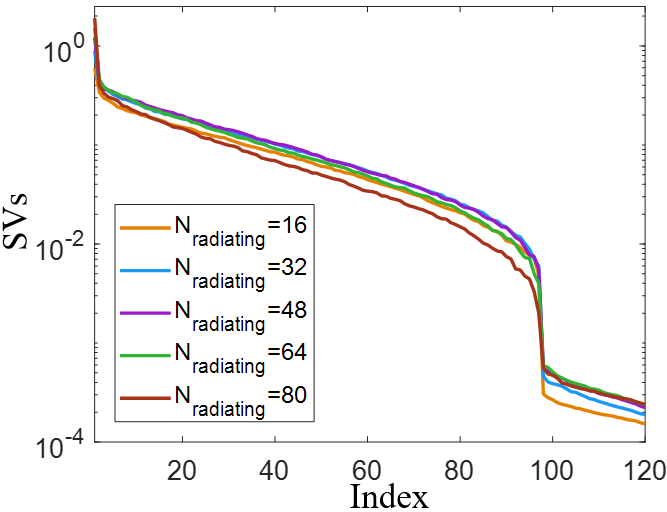}
	\caption[SVS: Varying $N_{radiating}$]{\label{fig:PC_svd_f1_nrad}The behavior of the SVS when changing the ratio of \emph{on}/\emph{off} elements.}
\end{figure}

In the above analyses we have turned on $50\%$ of the elements at a time. When more elements are radiating, the radiated power and consequently SNR will be higher, but the cavity's energy will also be depleted faster (reducing the contributions of elements further from the feed). Additionally, between a collection of masks, the patterns are more likely to have high correlation if many elements are \emph{on} since it is more likely that the masks will share a certain subset of $N_{radiating}$ elements. To assess the interplay between these factors, we vary $N_{radiating}$ (equivalently changing the \emph{on}/\emph{off} ratio) and plot the SVS. As can be seen in Fig. \ref{fig:PC_svd_f1_nrad}, the cases with $N_{radiating}=16$ and $N_{radiating}=80$ do not perform as well because of SNR and correlation considerations, respectively. A more moderate approach, such as the 50\% ratio, seems to perform the best. A similar trend was also observed in the simulated results with the analytic model of \cite{imani2018two}. It is worth noting that in \cite{imani2018two} SVS were normalized because the analytic model could not accurately account for (ohmic and radiative) losses. In contrast, here we have plotted unnormalized SVS where the role of SNR is more evident. The optimal performance of $N_{radiating}=48$ in this plot is the basis for our use of $50\%$ \emph{on}/\emph{off} throughout this paper (unless stated otherwise).

\subsubsection{Role of non-tunable elements\label{sec:nontunable}}

In the above analyses we came to the conclusion that the number of elements, $N_e$, acts as a limit on the number of useful measurements because the radiated patterns are a superposition of the contributions from the independent elements. However, this raises an interesting question: if we include additional elements that radiate but are not tunable, would we expect this relationship to change? In other words, since tunable elements are more costly to implement, can we increase the number of useful measurements by simply including non-tunable elements?

\begin{figure}[!t]
	\centering
	\includegraphics[width=0.8\columnwidth]{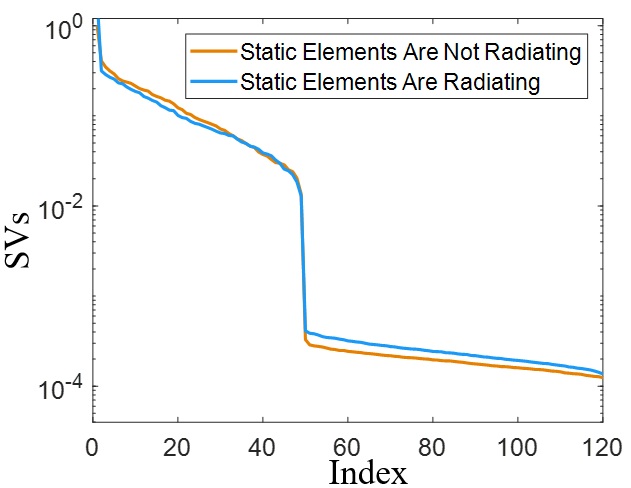}
	\caption[SVS: Static Elements]{\label{fig:PC_svd_f1_static}Comparison of when a fixed subset of $N_s=48$ elements are locked into a radiating or non-radiating state.}
\end{figure}

To answer this question, we have emulated a similar scenario using our fabricated DMA. Obviously we cannot include new static elements to probe this question, but we can take a subset of our existing elements, $N_{e,s}$, and lock them into an \emph{on} or \emph{off} state to effectively render them static. We therefore consider a case where the number of tunable elements is set to $N_{e,t}=48$ with $50\%$ of the elements \emph{on}, and the remaining $N_{e,s}$ are locked into either the radiating or non-radiating state. When the elements are locked into a non-radiating state it is as if the DMA only contains $48$ tunable elements, so we expect the SVS to fall off at $48$. For the static radiating case, the question is whether these elements add new information. From Fig. \ref{fig:PC_svd_f1_static}, being in an \emph{on} state does not appear to help the aperture obtain additional information.

This study answers a fundamental question about the intricate interplay between the cavity and the metamaterial elements, as discussed in \cite{marks2018tunablecavity}. From one perspective, we might expect that tuning the cELC elements serves to tune the modes inside the cavity. This makes sense since the elements leak energy out of the cavity eigenmodes, effectively changing the coefficients of the modes. However, this rebalancing of the coefficients does not have the effect of changing the accessible radiation patterns since the elements can only couple to the same modes that excited them. In order to create new cavity modes, the tunable components must support multiple resonating modes themselves. That is, the tunable components must be electrically large. Since this condition is not met for our subwavelength tunable elements, we see no enhancement from the inclusion of static elements. Our experimental result is therefore in alignment with the theoretical expectation detailed in \cite{marks2018tunablecavity}, where a more thorough discussion of these ideas can be found.

\begin{figure*}[!t]
	\centering
	\includegraphics[width=1\textwidth]{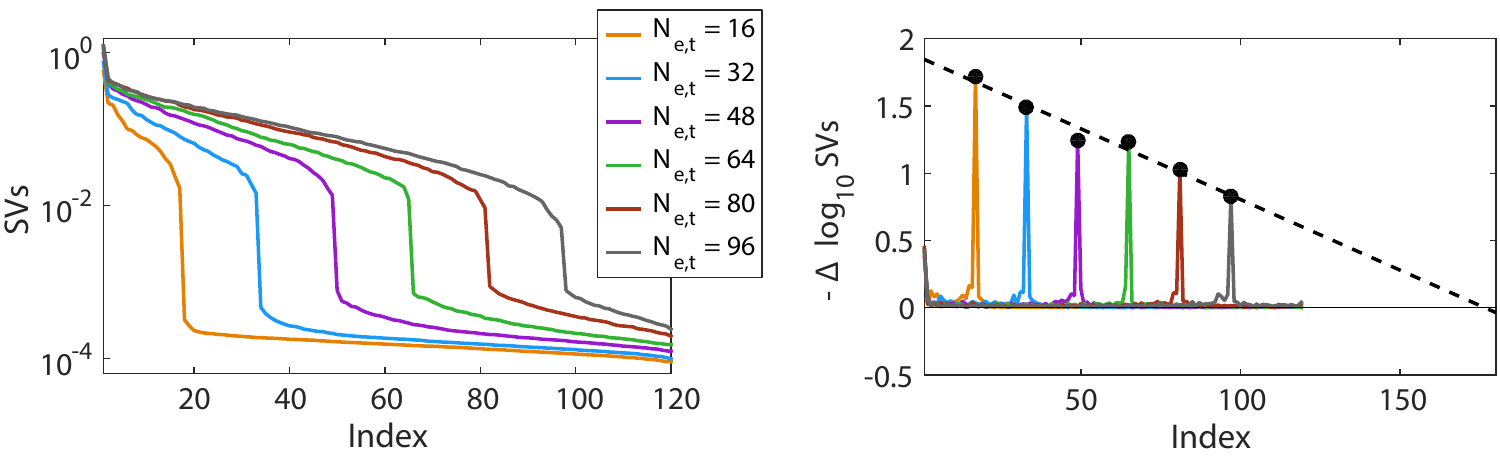}
	\caption[SVS: Varying $N_e$]{\label{fig:svd_Ne_ddx_combo}A study of how the SVS evolves when more elements are (effectively) utilized---static elements are fixed in an \emph{off} state. We also plot the derivative of the SVS and extrapolate the size/location of the drop off.}
\end{figure*}

\subsubsection{Number of cELC elements\label{sec:numelements}}

As described in \cite{Marks2016} and \cite{yurduseven2016design} and noted above, the number of elements in a multiplexing aperture is a key parameter. The number of elements sets the limit on the number of distinct measurements, while following a rigid trade off between SNR (when too few are used) and high level of correlation (when too many elements are used). In our structure, we have only used $96$ elements based on limitations of a practical PCB design. However, an important question remains: how much field diversity can we gain if we use a larger number of elements? To answer this question, we note that the experimental sample provides us with a unique opportunity to establish a trend between the number of elements and the quality of modes. This trend can be used to estimate the potential improvement in number of measurements given more aperture element. 

To deduce this trend, we start by examining the variation in the performance as we change the number of elements $N_e$. This can be accomplished, in effect, by locking a subset of the elements into a static \emph{off} state, and only tuning the remaining $N_{e,t}$ elements. In Fig. \ref{fig:svd_Ne_ddx_combo} we have completed this study for varying $N_{e,t}$ with each case having $50\%$ of the tunable elements in the \emph{on} state. The first observation from this plot is that the number of independent measurements is limited by $N_{e,t}$ as we would expect from previous studies. With increasing $N_{e,t}$ the SVS are enhanced, but with diminishing returns. From this plot we can attempt to extrapolate at what number of elements the limit of information content is reached. Specifically, when the sharp fall-off disappears, we would expect the information content of the system to be fairly close to saturation. We have therefore also plotted the derivative of the SVS (in log). The magnitude of the derivative of the fall-off shrinks as we increase the number of elements, and it can be seen that this trend follows a relatively linear relationship. By extrapolating this trend we see that the maximum of the derivatives ($\Delta$) of the SVS hits $0$ when $N_{e,t} \approx 177$. Therefore, if $N_e=177$ elements were included in the DMA we would anticipate the sharp cut-off to vanish. We would expect marginal improvements beyond this point, with the SVS still getting better but at a slower rate due to higher (radiation) losses. Of course, there is also the possibility that the behavior changes as more elements are included and that the relationship begins to converge asymptotically instead of linearly.

For an additional comparison, there is also the limit imposed by the space-bandwidth product, which is based purely on the aperture size and the wavelength \cite{Lohmann1996}. With independent radiating elements occupying the same aperture area, we know that the number of useful measurements is limited by $\text{SBP}_\text{aperture}$. Based purely on this geometry there is a hard upper bound at approximately $400$ which is twice larger than the estimate reached above. In reality it is known that this theoretical upper bound of $400$ is an overestimate because the resolution of the system falls off at wide angles, so the result we obtain could be closer to the true limit. Clearly there will still be some room for improvement beyond $N_e=177$, but such gains are difficult to reach without increasing the system complexity. It thus seems that this is a reasonable location in the design space to have a high-performance aperture with reasonable cost/complexity. Though we only included $96$ elements in our aperture, a more sophisticated biasing scheme would certainly make the inclusion of more elements possible, and future samples may enjoy more measurement modes by including more elements. Such a sample can be used to experimentally assess the proposals made in this subsection, and is left for future works.

\subsubsection{Grouping of the elements}

We were able to conduct the studies above in part because we have independent control over each metamaterial element. Generally speaking, most dynamic metasurfaces tie a large collection of their elements together in parallel and tune them as a group \cite{sievenpiper2002tunable,Lim2004,Gregoire2014,guzman2012electronic,Gregoire2015,fu2017electrically,chen2019electronically}. Such a scheme significantly simplifies the biasing configuration, and could be a way to increase the number of elements without dramatically increasing the system complexity. We can test this proposal by partitioning our $96$ elements into subgroups and actuating the subgroups in parallel. For example, we can split the $96$ elements into $24$ subgroups each containing $96/24=4$ elements. Using a single control \emph{knob} for each subgroup, we can then turn all elements in that subgroup \emph{on} or \emph{off}.

\begin{figure}[!t]
	\centering
	\includegraphics[width=\columnwidth]{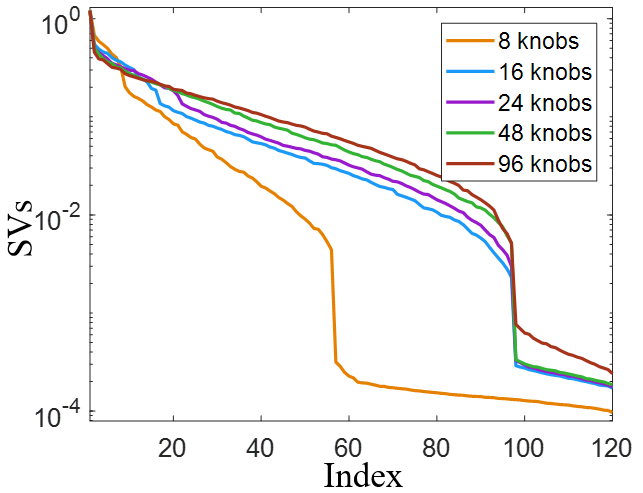}
	\caption[SVS: Parallel Control Lines]{\label{fig:PC_svd_f1_nknobs}Study to see how the SVS change when subgroups of elements are tied together and actuated in parallel.}
\end{figure}

We have investigated this tuning scheme and the results are plotted in Fig. \ref{fig:PC_svd_f1_nknobs}. The performance using $96$ knobs (no parallelization) exceeds all other cases, with ones such as $24$ and $48$ knobs not being substantially worse. As can been seen, dropping the number of knobs by too much can be detrimental. In fact, another interesting trend can be seen in the results with fewer knobs. For the cases of $8$, $16$, and $24$ knobs, there are \emph{kinks} in the SVS at index numbers $8$, $16$, and $24$, making multiple \emph{plateaus} in the plots. The first plateau (before the kink) can be thought of as first-order effects which are chiefly generated by the actuation of a single knob at a time. The following plateau is a second-order effect, where the mode has interacted and been scrambled into by a combination of two of the \textit{knobs}. This trend is in line with the theoretical prediction of \cite{marks2018tunablecavity}. Note that when we have grouped several elements into one effective tunable element, we then have electrically larger element and thus we see the second order effects as well (in contrast to the results of Subsection \ref{sec:nontunable}). While there are some trends between the theory in \cite{marks2018tunablecavity} and the empirical data here, the numerical results are not in perfect agreement. Nonetheless, both works suggest some underlying relationship among the modulators inside a cavity and the number of useful measurements. While there is still work to be done in this area, the results do suggest that a large number of elements can be tuned from a smaller number of control knobs so long as the elements are aggregated into relatively small subgroups. This gives an alternative path to increase the number of elements to the range of $200$--$400$ (cf. \ref{sec:numelements}).

\section{Imaging System Demonstration}
Previous sections presented an in-depth investigation into the multiplexing capabilities of the printed cavity DMA. We experimentally established the roles of different parameters on the 2D DMA performance. Since the ultimate promise of this work is computational imaging, we now utilize the 2D DMA in a simple imaging configuration and verify its ability to retrieve images of simple scene. 

\begin{figure}[!t]
	\centering
	\includegraphics[width=\columnwidth]{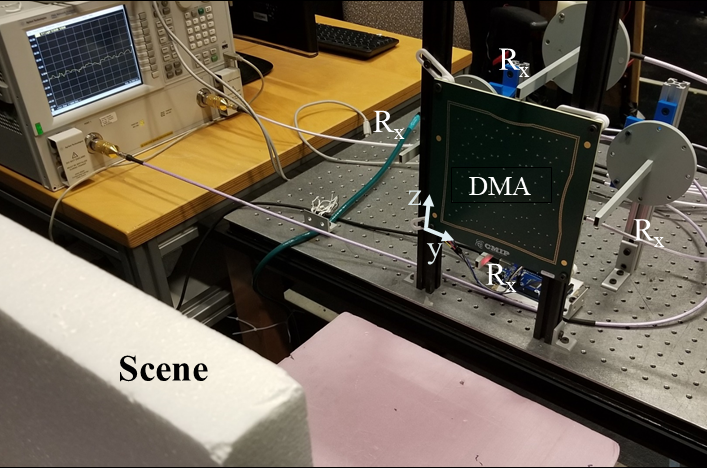}
	\caption{\label{fig:imagingsetup}Imaging setup with a DMA as the transmitter and four receiving open-ended waveguides.}
\end{figure}

As shown in Fig. \ref{fig:imagingsetup}, we use one DMA as the transmitting aperture and 4 open-ended waveguide probes as receivers. In this configuration, the probes have minimal spatial resolution and retrieving cross range resolution is primarily due to the 2D DMA. The DMA is connected to the port 1 of a vector network analyzer while port 2 is connected to the four probes via a mechanical switch. The measurement index in this setup thus loops through the masks, frequencies, and the receiving probes (cf. equation (\ref{eq:main})).

\begin{figure}[t]
\subfigure[Frequency-diverse aperture]{
	\centering
	\includegraphics[width=\columnwidth]{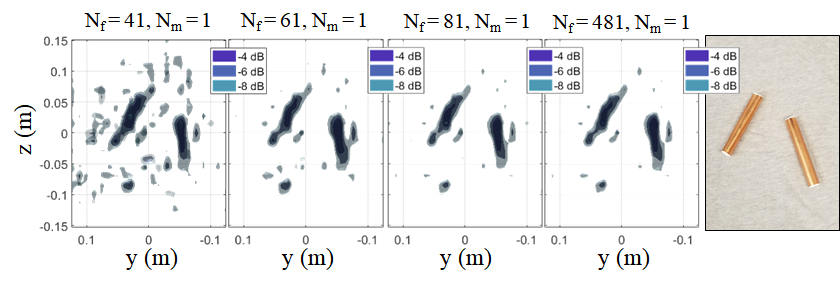}}
	\subfigure[Dynamic aperture]{
	\centering
	\includegraphics[width=\columnwidth]{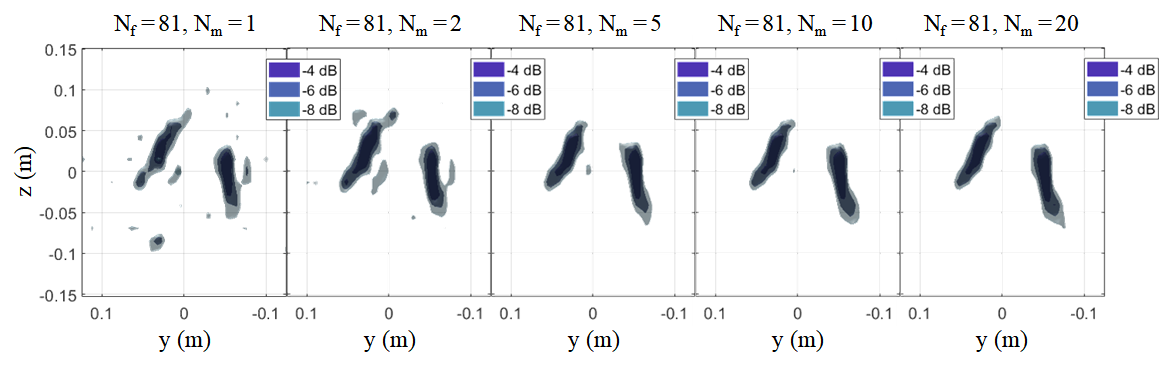}}
	
	\caption{\label{fig:imaging_passive_freq_pts} a) Imaging results as we increase the number of frequency points. b)Imaging results as we increase the number of masks while the number of frequency points is kept fixed.}
\end{figure}

The first imaging test is to assess the results of Fig. \ref{fig:svd_fm_combo} and illustrate the effects of the number of frequency points on the imaging performance. The reconstructed image and the target under test are shown in Fig. \ref{fig:imaging_passive_freq_pts}a. As we increase the number of frequency points we see improvement. However, this improvement, as predicted by Fig. \ref{fig:svd_fm_combo}, is only up until $80$ frequency points. Beyond this point we do not see much tangible improvement in the imaging performance.

Next, we keep the number of frequency points fixed at $N_f=81$, and increase the number of tuning masks, similar to the case study of Fig. \ref{fig:svd_fm_combo}b. The reconstructed images for this case are shown in Fig. \ref{fig:imaging_passive_freq_pts}b. We clearly see that increasing the number of masks in fact does improve imaging performance, with the final result significantly improved over the passive case. 

\begin{figure}[!t]
	\centering
	\includegraphics[width=\columnwidth]{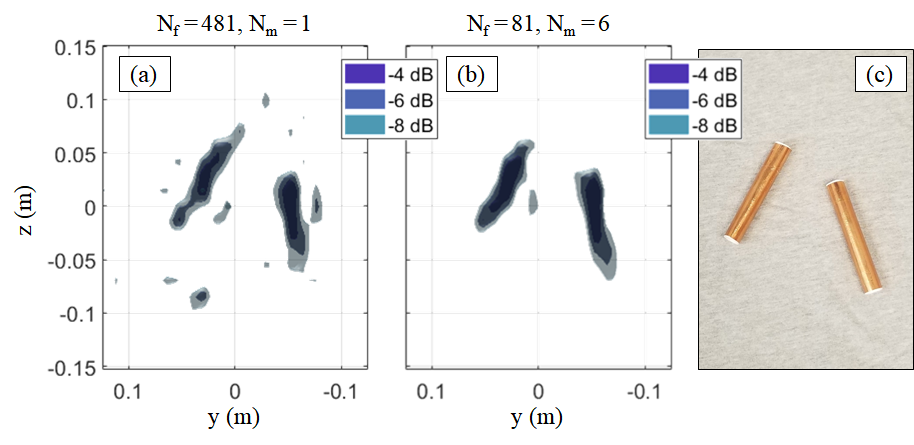}
	\caption{\label{fig:imagingexample1}Imaging results a) without using tuning masks of the DMA b) with using tuning masks of the DMA. c) target under test.}
\end{figure}

A more fair comparison of frequency-diverse versus dynamic aperture imaging performance can be conducted by using similar setting as in Fig. \ref{fig:PC_svd_f1_nmasks}. More specifically, we use the DMA with the same number of measurements to image the object; once using only frequency diversity with $N_f=481$ frequency points and another time using $N_f=81$ frequency points and $N_m=6$ tuning masks. The result for both cases in Fig. \ref{fig:imagingexample1}. We clearly see a higher quality image when using the tuning masks, in line with the predictions of Fig. \ref{fig:PC_svd_f1_nmasks}. 

\section{Conclusion}

This paper described a fabricated 2D dynamic aperture and used that in simple experiments to obtain a detailed perspective into its operation and physics. Our focus was examining this device's ability to generate complex patterns that can multiplex a scene information, and identify factors that impact this capability. We then used the 2D DMA in a simple imaging system to validate the predictions about its performance in imaging context. The promising results presented in this paper suggest that a 2D DMA can operate as the central hardware of a computational microwave imaging system. 

To extend the present results, a larger imaging system with multiple DMAs can be used. Such a system could involve multiplexing both on the transmitting and receiving aperture, promising much higher quality imaging than those demonstrated to date at microwave frequencies \cite{Gollub2017}. Such a system can also pave the way for taking advantages of unique capabilities a 2D DMA can bring: for example, single frequency imaging \cite{sleasman2017single,boyarsky2018single}, as well as computational polarimetric imaging \cite{fromenteze2017computationalA}. This DMA can also be used in other contexts such as computational sensing \cite{del2018dynamic,del2018precise} or communication systems \cite{yoo2019enhancing,shlezinger2019dynamic}. Having established the promising multiplexing capability of the 2D DMA, another future direction can be to sculpt the complex patterns, for example to match the requirements of a feature-specific imaging system \cite{Neifeld2003,Liang2015} or to co-design the patterns and the classification algorithms \cite{chakrabarti2016learning,horstmeyer2017convolutional}. Toward this goal, we can leverage recent progress in analytical modeling of such complex structures using coupled dipoles \cite{pulido2016AWPL,pulido2017polarizability,pulido2018analytical,YooImani2019}.

\section{Acknowledgement}

This work was supported by the Air Force Office of Scientific Research (AFOSR) (FA9550-12-1-0491 and FA9550-18-1-0187).

\bibliographystyle{IEEEtran}
\bibliography{sample}


\end{document}